\documentclass[prl,twocolumn,aps,showpacs,amsmath,amssymb]{revtex4-2}

\usepackage{graphicx}
\usepackage{epstopdf}
\usepackage{dcolumn}
\usepackage{bm}
\usepackage{amsmath}
\usepackage{easyReview}

\usepackage{amsmath} 

\usepackage{graphicx}
\usepackage{amsmath,amssymb,amsfonts}
\usepackage{textcomp}
\usepackage{hyperref}
\usepackage{gensymb}
\usepackage{verbatim}
\usepackage{braket}

\hypersetup{breaklinks=true,colorlinks=true,urlcolor=black}
\usepackage{color}
\usepackage{soul}
\usepackage{gensymb}

\begin{document}

	\title{Unveiling Symmetry Instability induced by Topological Phase Transitions}
	
	\author
	{Liang Luo$^{1\dag}$, Boqun Song$^{1,2\dag}$, Genda Gu$^{3}$, 
		Martin Mootz$^{1}$, Yongxin Yao$^{1}$, Ilias E. Perakis$^{4}$, Qiang Li$^{3,5}$ and Jigang Wang$^{1,2^\ast}$}

	\affiliation{$^1$Ames National Laboratory, Ames, IA 50011, USA.
		\\ $^2$Department of Physics and Astronomy, Iowa State University, Ames, Iowa 50011, USA
		\\$^3$ Condensed Matter Physics and Materials Sciences Department, Brookhaven National Laboratory, Upton, NY 11973-5000, USA.
		\\$^4$ Department of Physics, University of Alabama at Birmingham, Birmingham, AL 35294-1170, USA
		\\$^5$ Department of Physics and Astronomy, Stony Brook University, Stony Brook, New York 11794-3800, USA
		\\ $^\ast$Correspondence: jgwang@ameslab.gov
		\\$^\dag$Equal contributions}

	\date{\today}

	\begin{abstract}
		The symmetry-topology interplay dictates how to define order parameters and classify material ordered phases. 
		However, current understanding of this interplay has been predominately approached from a one-sided perspective, with topological states being classified within the constraints imposed by specific fixed symmetries. 
		Here we complete this full circle by demonstrating spontaneous symmetry breaking that results from a periodic alteration of topological phases induced by light in a centrosymmetric Dirac material ZrTe$_5$. The distinguishing feature is the observation of robust correlation and striking anomalies in the fluence and temperature dependence of key transport parameters.
	First, both shift current $J_{\text{s}}$ and displacement current $J_{\text{d}}$, arising from interband transition and infrared phonon driving, respectively, along with charge carrier pumping, exhibit similar behaviors. Second, they all peak at similar low pump fluence, followed by a subsequent reduction as the fluence further increases. This behavior cannot be explained by conventional energetically allowed, direct excitations. Third, all the three observables exhibit anomalies when they approach the topological phase transition temperature. These results highlight the unique low-energy pumping behaviors in ZrTe$_5$, characterized by reversible fluence dependence and a 'hinge-like' interaction that connects various electronic and lattice observables, including phonons, charge carriers, and currents.
Our findings, supported by model analysis, provide key insights into the fragility of crystalline (inversion) and time-reversal symmetries during the dynamics of topological phase transitions. This fragility drives spontaneous symmetry breaking, evidenced by the synchronized emergence of off-resonant infrared phonons and broken-symmetry photocurrents. 
	\end{abstract}

\maketitle

\section{I. INTRODUCTION}

Symmetry is at the heart of the Landau-Ginzburg-Wilson scheme for classifying matter phases, and a phase transition is accompanied with spontaneous symmetry breaking (SSB) \cite{1,2,3,4,5}. Topology provides a beyond-symmetry classification and reveals a class of matter states, which are symmetrically identical but topologically distinct. 
Recently, there has been a growing recognition of the intricate interplay between symmetry and topology \cite{Ref1,Ref2}. This interplay serves as a pervasive theme in the fields of topological matter, search for new quantum materials, control of quantum materials properties \cite{Ref20}, and topological photonics \cite{Ref21}. Their development enables the study of materials through interactions with photons and phonons \cite{Ref22,Ref23,Ref24,Ref25,Ref26,j1,j2,j3, 1}.
Remarkably, in all of these scenarios, symmetries consistently serve as ``preconditions" under which topology is subsequently explored. In other words, symmetry acts as a ``premise," while topology emerges as an ``aftermath." 
Hence, a compelling, yet still elusive question remains: Is the interplay between symmetry and topology mutual? 
In other words, can a pristine symmetry experience instability and spontaneous disruption due to the influence of a driving topological phase transition (TPT)?

One might question how SSB could possibly be involved, because topology is usually used to classify states that share identical symmetries \cite{Ref1,Ref2,Ref3,Ref4,Ref5,Ref6,Ref8,Ref9,Ref10,Ref11,Ref12,Ref13,Ref14,Ref15,Ref16,Ref17,Ref18, Ref19,Ref20,Ref27}.  Indeed, within the context of equilibrium states, SSB appears irrelevant. However, when considering non-equilibrium processes, the question becomes more pertinent: when a TPT is happening, is there a transient period in which broken symmetries take place? In other words, can a topological state transition transiently disturb its pristine symmetry? It is important to note that both topological states and symmetries are conventionally defined under equilibrium conditions, which pertain to a long-time asymptotic behavior. Thus, this conventional framework does not preclude the occurrence of SSB during transient processes. Whether a SSB occurs or not, it does not contradict the currently established  framework of topology classification. Instead, it provides deeper insights into the dynamic processes involved in connecting two topological states. 
 
In fact, an argument valid in equilibrium may not necessarily apply to non-equilibrium scenarios. Thus, it is essential not to uncritically assume a constant pristine symmetry beforehand, especially for ultrafast processes with femtosecond time resolutions.
Recently, we observed that infrared (IR) phonon modes associated with the IS breaking can be generated, even far from resonance, in centrosymmetric materials \cite{Ref26}. The generation of these IR phonons, typically prohibited by the pristine symmetry of the material, takes place in close proximity to a topological phase transition. This observation alludes to certain unknown mechanisms and raises an intriguing perspective: whether the IS breaking by IR phonons has a topological origin. 

In this paper, we examine the response of symmetry breaking, specifically IS and time-reversal symmetry (TRS), during a switchable TPT in the centrosymmetric Dirac material ZrTe$_5$. The intensity of IS (TRS) breaking is characterized by the magnitude of displacement current $J_{\text{d}}$ (shift current $J_{\text{s}}$), associated with an IR phonon mode (interband charge pumping). Briefly, we find that both the displacement current $J_{\text{d}}$ and shift current $J_{\text{s}}$ are enhanced (or suppressed) when TPT is turned on (or off). We attribute these observations specifically found in ZrTe$_5$ as preliminary clues for a generic principle for symmetry-topology interplay. We have also conducted cross-checks by comparing fluence- and temperature-dependent results and performing model analysis to ensure that our observations are indeed associated with TPT and not merely coincidence. 
Our experiments, supported by model analysis, reveal an intriguing physics scenario that periodic modulation of TPTs induced by laser-excited phonons can lead to the fragility of time-reversal and inversion symmetries, as manifested by the generation of IS breaking IR phonons (displacement current $J_{\text{d}}$) and TRS breaking currents $J_{\text{s}}$. To differentiate the initial phonons (any phonons, which can be excited by laser pulses) for triggering TPTs and the subsequent IS breaking IR phonons, we refer the former as ``phonons" and the latter as ``IR phonons" or ``B$_\text{1u}$ phonons" throughout this article. 
Moreover, the fluence- and temperature-dependent correlations between different observables (e.g., currents, charge pumping), as well as their striking anomalies also help rule out other scenarios.

\begin{figure*}
	\begin{center}
		\includegraphics[width=160mm]{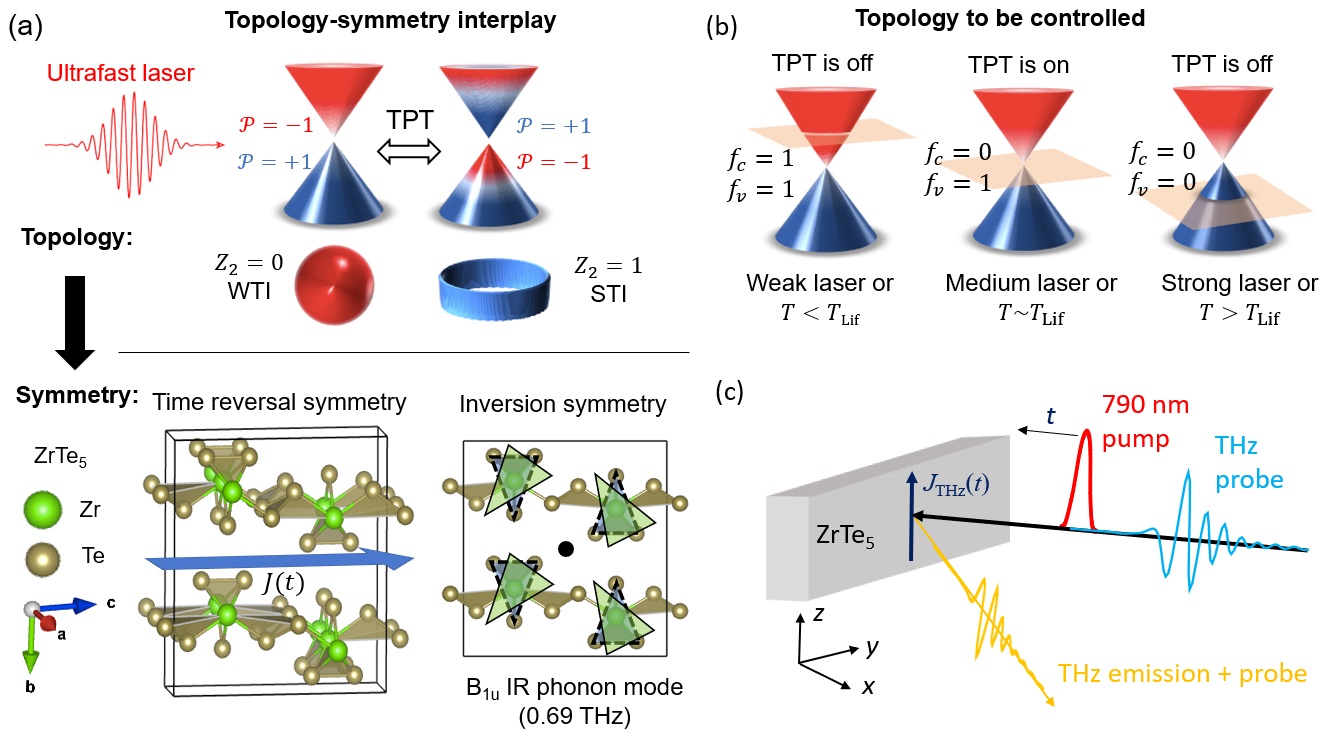}
		\caption{(a) A schematic illustrating the topology and symmetry interplay. An ultrafast laser pulse induces phonons, which lead to band evolution and cause periodic TPT (top panel). For ZrTe$_5$, TPT is between STI and WTI. Topological transition will cause symmetry breaking, including time-reversal symmetry and inversion symmetry (bottom panel). The photocurrent $J(t)$ arises from time-reversal symmetry breaking (bottom left panel) and the B$_\text{1u}$ IR phonon mode arises from inversion symmetry breaking (bottom right panel, in which the black dot indicates the inversion center). Both $J(t)$ and the B$_\text{1u}$ IR phonons will emit radiation of THz frequency, which is measured using THz emission spectroscopy. (b) The control of TPT by tuning the chemical potential $\mu$ (orange plane) with either laser pulse (weak vs. medium vs. strong excitations) or temperature $T$. A laser pulse will cause a downshift of $\mu$ transiently with respect to the Dirac point (see text). Temperature will modify the lattice parameter, leading to the shift of $\mu$ as well. When $\mu$ crosses the cone vertex, namely at the Lifshitz transition, a TPT is turned on (middle panel). When $\mu$ is away from the cone vertex, the TPT is turned off (left and right panels). (c) Our experimental scheme combines optical pump-THz probe spectroscopy and THz emission spectroscopy. Two kinds of measurements are performed: the THz probe is either on or off. When it is off, the THz emission signal is exclusively measured from the photocurrent $J(t)$ and B$_\text{1u}$ IR phonon mode. When it is on, charge carrier dynamics is measured in addition to the photocurrent and IR phonon mode.}
		\label{Fig1}
	\end{center}
\end{figure*}

\section{II. Experimental Scheme and Materials}
ZrTe$_5$ is selected for investigating the correlation between TPT and breaking of symmetries: specifically IS and TRS (Fig. 1(a)). In the pristine state, ZrTe$_5$ preserves both symmetries and lies in proximity to a topological transition between strong (STI) and weak (WTI) topological insulators \cite{Ref28,Ref29,Ref30,Ref31}. 
ZrTe$_5$ has a single band cone at the $\Gamma$ point with a bandgap of $\sim$40--100 meV that varies with samples \cite{Ref32,Ref33}. For our sample, it exhibits a STI phase (1;110) at low temperature and transits to a WTI phase (0;110) above the Berry temperature $T_\text{Berry}\sim$150 K due to the thermal change of lattice parameters \cite{Ref25,Ref26}. 
The chemical potential $\mu$ lies above the Dirac cone vertex at low temperature $T$=5 K and decreases downwards with increasing temperature. At the Lifshitz transition temperature $T_\text{Lif}\sim$60 K, $\mu$ crosses the Dirac cone vertex. These are confirmed by both our previous THz emission measurement \cite{Ref26} as well as by transport measurements \cite{Ref28}. More info about the sample can be found in the Supplemental material. 

A crucial component in experimental design is a ``switch" for TPTs. In general, the topological invariant $\upsilon$ is defined as the product of parities ${\zeta}_i$ for all the occupied bands at time reversal invariant points \cite{Ref1,Ref34,Ref35}. Since ZrTe$_5$ features a single band cone, its topological index $\upsilon$ boils down to parities at $\Gamma$ point in the Brillouin zone (BZ). 
\begin{eqnarray}  \label{equ1}
\begin{aligned}
(-1)^{\upsilon}=\prod_i^N {\zeta}_{i}.
\end{aligned}
\end{eqnarray}
where ${\zeta}_{i}$ is the $i$-th band parity at $\Gamma$. ${\zeta}_{i}=1$ or $-1$, making a $Z_2$ index ${\upsilon}=0$ or $1$. The $N$ is the total number of occupied bands. ZrTe$_5$ features valence and conduction bands with opposite parities, i.e., ${\zeta}_N=-{\zeta}_{N+1}$, and band inversions (${\zeta}_N{\leftrightarrow}{\zeta}_{N+1}$) lead to a switch of $\upsilon$ between 1 and 0. However, if both the conduction and valence bands become empty, ${\zeta}_N$ and ${\zeta}_{N+1}$ will be excluded from the product in Eq. (1), and their inversion will cause no effect on $\upsilon$, i.e., TPT is turned off. Here we employ laser pulses and temperature to tune $\mu$ to realize this mechanism. That is, when $\mu$ crosses the cone vertex, TPT is turned on; when $\mu$ moves away from the cone vertex at $\Gamma$ (both bands are empty), TPT is turned off. The virtue of this approach is that one does not need to precisely determine the critical point for $\mu$ to pass the vertex, but simply performs a fluence or temperature dependent measurement to observe the trend as $\mu$ is shifted across the band cone. The mechanism is illustrated in Fig.~1(b).


Regarding the experiment setup shown in Fig.~1(c), the sample $a$ and $c$ axes are along $y$ and $z$ axis, respectively. We use optical pump with a central wavelength of $\sim$790 nm (1.57 eV) and pulse duration $\sim$40 fs, which is incident on the sample at 45 degree. In addition, we add a THz probe polarized along $z$ axis to measure charge carrier dynamics. Both THz probe and THz emission are measured along $z$ axis and in reflection geometry, as the majority of THz emission is along the sample IS breaking $c$-axis \cite{Ref26}. This way, we can simultaneously monitor the evolution of photocurrents and charge carriers subject to multi-dimensional parameter tunings that include electric field strengths, phonon amplitudes, and, most importantly, TPTs. 

Applied to ZrTe$_5$, the extent of TRS and IS breaking illustrated in Fig.~1(a) are characterized by two observables: the amplitude of the TRS-breaking photocurrents (blue curve in Fig.~2(a)) and the amplitude of the IS-breaking IR phonon mode (orange curve in Fig.~2(a)). Later, we will show that the two observables originate from shift current $J_{\text{s}}$ and displacement current $J_{\text{d}}$, respectively. These transient currents emit radiation in THz frequencies, which are measured by THz emission spectroscopy. Moreover, charge carrier dynamics (black curve in Fig. 2(a)) are measured by optical pump-THz probe spectroscopy. Details of the experimental setup can be found in Ref \cite{Ref26}.

\begin{figure}
	\begin{center}
		\includegraphics[width=80mm]{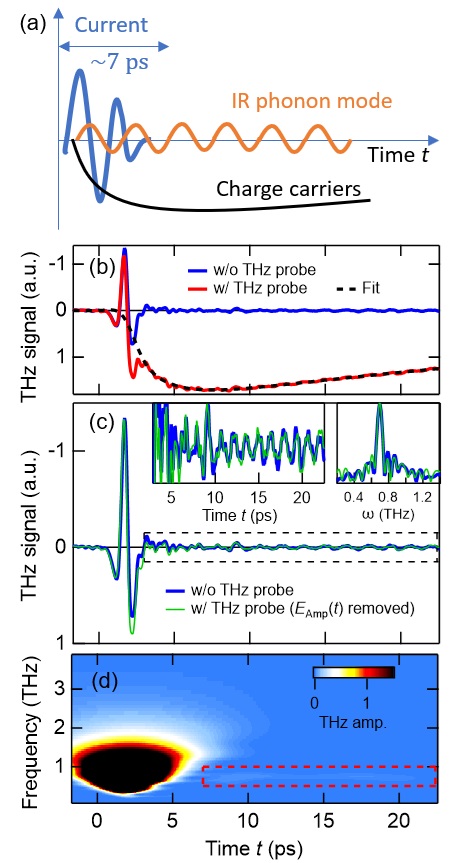}
		\caption{Representative pump-induced THz signals at $T$=5 K.  (a)  A simplified schematic (not to scale) illustrates the three components of the pump-induced  THz signals measured (see text): photocurrent (blue curve), IR phonon mode (orange curve), and charge carrier pumping (black curve).
			(b) Pump-induced THz emission signals as a function of time, with (red curve) and without (blue curve) the THz probe. The black dashed line is the corresponding fitting by the single exponential rise and decay equation (Eq. 2 in the text).(c) Comparison of the THz signals in (b) with the amplitude profile (black dashed line in (b)) removed. The left inset shows the close-up of the THz signals in the dashed box, which exhibit a clear THz periodic oscillatory signal after $\sim$7 ps. The right inset shows the corresponding Fourier Transform spectra for $t>$7 ps. (d) Wavelet analysis of the blue curve in (b) shows two features: first, a broadband THz emission signal, which ranges from $\sim$0.2-3 THz and lasts for $\sim$7 ps, arises from the photocurrent; second, a narrow peak which is centered at $\sim$0.69 THz and lasts for tens of ps, arises from the IR phonon mode.} 
		\label{Fig2}
	\end{center}
\end{figure}

\section{III. Experimental results}

Representative pump-induced THz signals at low temperature $T$=5 K are shown in Fig. 2(b). When the THz probe is off, the THz signal solely originates from pump-induced THz emission (blue curve), which exhibits a THz main peak followed by small periodic oscillations lasting for tens of ps. When the THz probe is on, the THz signal exhibits an additional amplitude profile (red curve), which can be fitted well by the single exponential rise and decay equation (black dashed curve),
\begin{equation} \label{eq:amp}
E_\text{Amp}(t)\sim A_\text{charge}(1-e^{-t/\tau_1})(e^{-t/\tau_2}), 
\end{equation}
where $t$ is the time and $A_\text{charge}$ is the amplitude. The fitted rise time constant is $\tau_1 \sim$2.5 ps, and the decay time constant is $\tau_2 \sim$40.6 ps. After removing the fitted amplitude profile $E_\text{Amp}(t)$ by Eq. 2, the THz signals with and without the THz probe match each other very well, as shown in Fig. 2(c). This indicates that the amplitude profile arises exclusively from the THz probe, which we attribute to the pump-induced charge carrier dynamics to be discussed later. 

We verify our extraction of data from different aspects. The left inset of Fig. 2(c) shows the close-up of the THz signals enclosed in the dashed box of Fig. 2(c), which exhibit a clear THz periodic oscillatory signal after $t\sim$7 ps. The right inset of Fig. 2(c) shows the corresponding Fourier transform (FT) spectra performed for $t>$7 ps, which exhibits a resonance centered at $\sim$0.69 THz. Note that the reason we chose $t>$7 ps for FT is to get rid of the coexisting signal from the THz main peak that persists mainly for $t<$7 ps. Fig. 2(d) shows the wavelet analysis of the blue curve from Figs. 2(b) and 2(c), which is characterized by a broadband THz emission peak ranging from $\sim$0.2-3 THz and lasting for $\sim$7 ps, and a narrowband peak centered at $\sim$0.69 THz lasting for tens of ps (highlighted in the dashed red box). 


Importantly, Figs. 2(b)-2(d) suggest three coexisting signals from the photoexcited ZrTe$_5$, which are summarized in a simplified schematic in Fig. 2(a). The first one is attributed to the pump-induced photocurrent $J$ (blue in Fig. 2(a)), which mostly lasts for $\sim$7 ps, as evidenced by the main THz emission peak in Fig 2(d). The second one is the periodic oscillation with a frequency of $\sim$0.69 THz (orange in Fig. 2(a)), which lasts for tens of ps and is attributed to the pump-induced B$_\text{1u}$ IR phonon mode \cite{Ref26}, as shown in Figs. 2(b)-2(d). In order to exclusively resolve the IR mode spectra, a FT has been performed for $t>$7 ps in order to get rid of the coexisting photocurrent spectra, which mainly exist for $t<$7 ps. We note that our measurements do not capture the complete IR phonon relaxation dynamics due to the limited scan range imposed by the echo signal from the 1 mm ZnTe detector, which appears $\sim$24 ps after the THz main peak. Based on the inset of Fig. 2(c), the decay time of this IR mode appears much longer than 22 ps as the oscillations do not exhibit any clear relaxation within this period of time. However, the decay time of this IR mode is non-topological and does not impact the focus of this study. The last component is the amplitude profile $E_\text{Amp}(t)$ following the single exponential rise and decay lineshape (black in Fig. 2(a)), which is attributed to photoexcited charge carrier dynamics.


\begin{figure}
	\begin{center}
		\includegraphics[width=90mm]{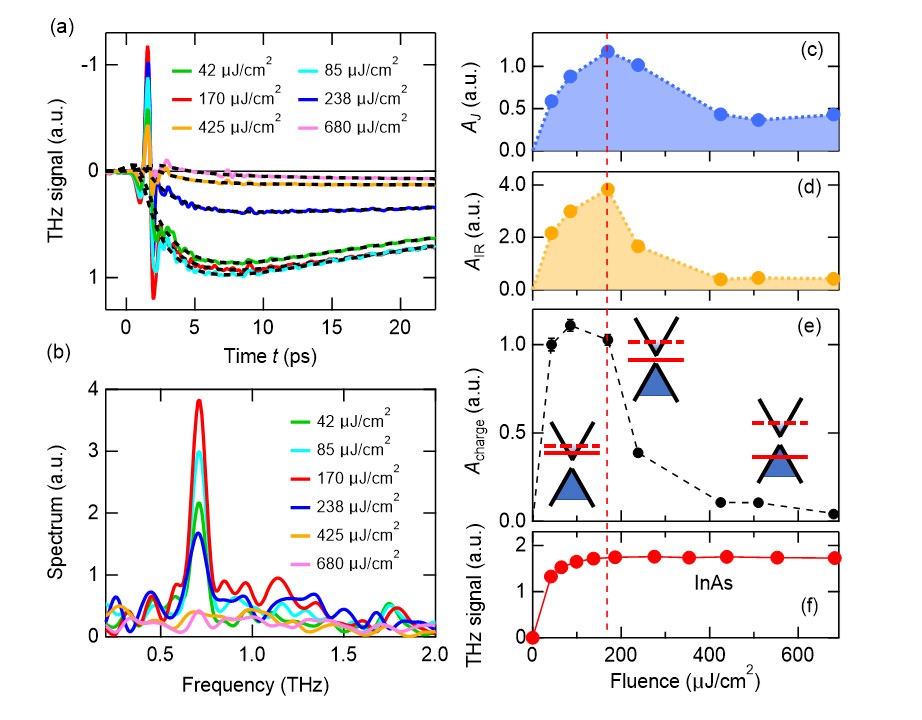}
		\caption{Fluence dependent THz signals at $T$=5 K. (a) Pump-induced THz signal as a function of time for several selected fluences. The black dashed lines are corresponding fittings by Eq. 2. (b) The FT spectra of the residual oscillatory signal from (a) for $t>$7 ps after removing the fitted amplitude profile (see text), which shows the IR phonon mode centered at $\sim$0.69 THz. (c) Fluence dependence of the photocurrent amplitude $A_J$ extracted from (a). (d) Fluence dependence of the IR phonon amplitude $A_\text{IR}$ extracted from (b). (e) Fluence dependence of the charge pumping amplitude $A_\text{charge}$ from fitting by Eq. 2. The inset shows bandstructure of ZrTe$_5$ following excitation with a low fluence (left), threshold fluence $F_\text{th} \sim$170 $\mu$J/cm$^2$ (middle), or high fluence (right), respectively. The dashed (or solid) horizontal red lines indicate static (or pump-induced transient) Fermi levels of ZrTe$_5$. The vertical red dashed line marks the position of fluence at $F_\text{th}$. (f) Fluence dependence of the photocurrent amplitude obtained by a similar THz emission measurement using the same setup from a topologically-trivial, narrow bandgap semiconductor InAs at room temperature. }
		\label{Fig2}
	\end{center}
\end{figure}

Next, we investigate the three observables at various pump fluences at $T$=5 K, as shown in Fig. 3. Pump-induced THz signals at a few selected fluences together with the corresponding amplitude profile fittings are shown in Fig. 3(a). After removing the fitted amplitude profile from the THz signals, the residual periodic oscillations for $t>$7 ps are Fourier-transformed in the same way as Figs. 2(b)-2(c). The IR phonon mode centered at $\sim$0.69 THz is clearly observed, as plotted in Fig. 3(b). 
The IR mode amplitude $A_\text{IR}$ as a function of fluence is plotted in Fig. 3(d). The magnitude of $A_\text{IR}$ determines the intensity of IS breaking \cite{Ref26}, therefore, Fig. 3(d) suggests that when the pump fluence increases above the threshold $F_\text{th} \sim$170 $\mu$J/cm$^2$, the extent of IS breaking is reduced. Fig. 3(c) plots the amplitude of the photocurrent, $A_J$, as a function of fluence, in which $A_J$ is taken as the amplitude of the THz main peak $E_\text{THz}$ in Fig. 3(a), because $A_J\sim E_\text{THz}$ \cite{Ref26,BraunNC2016}. The amplitude of charge pumping $A_\text{charge}$ is obtained from the amplitude profile fitting by Eq. 2 and plotted in Fig. 3(e). We note two remarkable observations. First, $A_J$ peaks at $F_\text{th}\sim$ 170 $\mu$J/cm$^2$ and decreases with further increasing fluence, which indicates that a higher fluence is unfavorable for photocurrent generation. This suggests that the photocurrent is $\textit{induced}$ by light rather than being directly $\textit{driven}$ by light. The former is defined as the photocurrent being indirectly caused by the presence of light, but not directly generated by light itself. The latter is defined as the photocurrent being created and actively maintained solely by the presence of light, and once the light field is removed, the photocurrent diminishes. Therefore, a key signature of light field-driven photocurrent is that $A_J$ monotonically increases with fluence and may eventually saturate at a very high fluence. The decrease of $A_J$ above $F_\text{th}$ observed in our measurements clearly indicates that the photocurrent $J$ is not directly driven but indirectly induced by light. Thus, the nonlinear response relation $J\sim E^2$ reported in other works \cite{Ref27,Ref36,Ref37} does not apply here \cite{Ref38}. Similarly, the fluence dependence of $A_\text{charge}$ shown in Fig. 3(e) cannot be interpreted by a direct and energetically resonant pumping mechanism, because no matter whether pumping occurs  via single- or multi-photon excitations, or photon-induced phonons, or other energetic quanta, $A_\text{charge}$ should increase with fluence until saturation. Second, remarkably, we notice a correlation among the three observables ($A_J$, $A_\text{IR}$, $A_\text{charge}$) which exhibit a similar fluence dependence, i.e., they all reach the maximum near $\sim F_\text{th}$ and decrease above it. These three distinct observables are related to photocurrents, IR phonons, and charge carrier population, and they all exhibit a strong correlation in their fluence dependence.  In addition, the striking decrease observed above the relatively low fluence $F_\text{th}$ in our measurements of ZrTe$_5$ is clearly distinct from a similar THz emission measurement of InAs (Fig. 3(f)) obtained using the same setup, thereby excluding technical issues or artifacts due to instruments. The observation on ZrTe$_5$ is also different from THz shift current emission from topologically-trivial semiconductor GaAs \cite{DCoteAPL2002}. These correlation and control measurements clearly indicate a common origin related to the topological phase transition in ZrTe$_5$. 



The abnormal pump fluence dependence of the three observables $A_J$, $A_\text{IR}$, and $A_\text{charge}$ are remarkable. They clearly show that they are “hinged”, i.e., favorable pumping conditions for generating photocurrents are also favorable for exciting IR phonons and charge carriers. Strikingly, the energy scales for generating $A_J$, $A_\text{IR}$, and $A_\text{charge}$ are quite distinct, from a few meV to eV range, as listed in Table I. Photocurrent generation arises from gapless excitations, because ZrTe$_5$ exhibits slight $n$-doping at low temperature 5 K, making it akin to a metallic system; IR phonon excitation with 0.69 THz resonance requires a photon with energy of $\sim$2.85 meV (=0.69 THz); Charge excitation requires to overcome the bandgap, which is about tens to hundreds of meV for ZrTe$_5$. Note that optical excitations arising from various energy scales typically require different optimal pumping conditions. Similar results for very different energy scales usually imply the absence of a characteristic energy \cite{Ref40,Ref41}, i.e., the system’s behavior is not characterized by an energetic parameter, but possibly by geometric/topological ones.


The chemical potential $\mu$ of ZrTe$_5$ is located above the Dirac point at $T$=5 K, because ZrTe$_5$ is inadvertently $n$-doped due to imperfect crystal growth \cite{Ref28}. When an ultrafast laser pulse is applied, two effects occurred: (1) charge excitation from valence band to conduction band, leading to increased electronic energy, and (2) increased electronic and lattice temperature, leading to an upward shift of the topological band structure. Both effects combined can result in relative downward shift of the Fermi level with respect to the Dirac band, as shown in time-resolved ARPES measurements of ZrTe$_5$ \cite{Manzoni2015}. Therefore, $\mu$ is transiently suppressed after optical excitation (inset of Fig. 3(e)). The low- and high-fluences may lead to $\mu$ being located transiently above and below the Dirac point (or Weyl point if IS breaking significantly splits the Dirac cone) respectively, i.e., ${\zeta}_i$ and ${\zeta}_{i+1}$ are both occupied or empty. Therefore, fluence dependence actually reflects $\mu$ dependence. We argue that the maximum amplitude of all three observables near $F_\text{th}$ corresponds to $\mu$ crossing the cone vertex, at which point the carrier type (electron or hole) is switched, namely at the Lifshitz transition \cite{Ref26,Ref28}. However, we note that the Lifshitz transition does not modify the band degeneracy (gap-closing) and thus it does not directly lead to TPT. TPT is caused by light-induced phonons (any phonons, which can be excited by the 790 nm laser pulses) \cite{Ref26}. At the Lifshitz transition, the occupancy drastically changes between valence and conduction bands (e.g. $i^\text{th}$ and $(i+1)^\text{th}$ bands), which has a profound effect on the inversion for ${\zeta}_i$ and ${\zeta}_{i+1}$ in Eq. (1). 

\begin{figure}[tbp]
	\begin{center}
		\includegraphics[width=90mm]{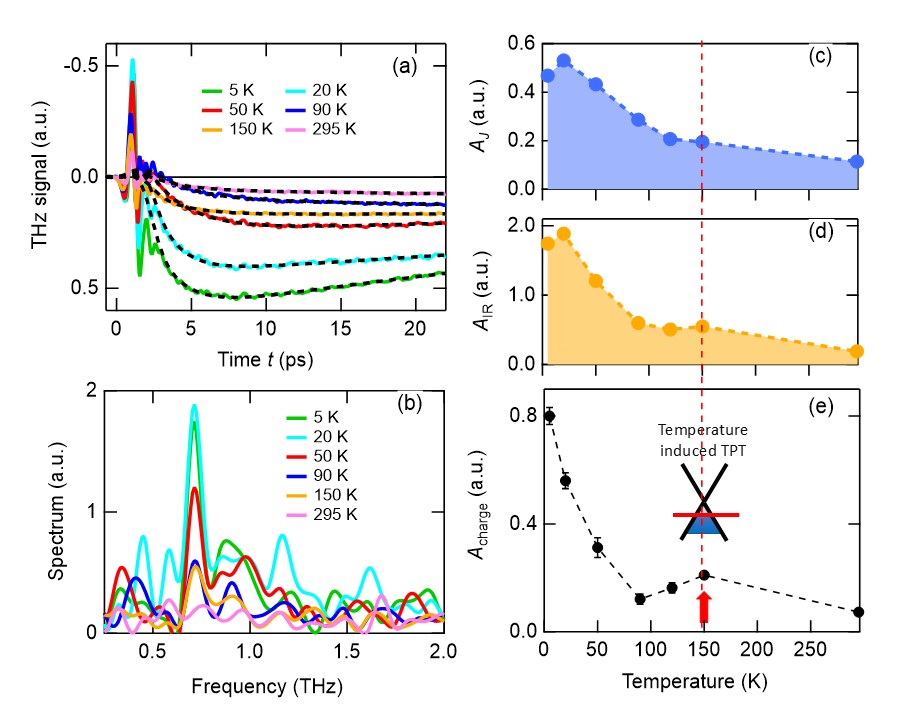}
		\caption{Temperature dependent THz signals at a pump fluence $F$=40 $\mu$J/cm$^2$. (a) Pump-induced THz signal as a function of time for several selected temperatures. The black dashed lines are corresponding fittings using Eq. 2. (b) The FT spectra of the residual oscillatory signal from (a) for $t>$7 ps after removing the fitted amplitude profile (see text), which shows the IR mode centered at $\sim$0.69 THz. (c) Temperature dependence of the charge pumping amplitude $A_\text{charge}$ from fitting using Eq. 2. Temperature dependence of the (d) IR phonon amplitude $A_\text{IR}$ and (e) photocurrent amplitude $A_J$ extracted from (b) and (a), respectively. }
		\label{Fig3}
	\end{center}
\end{figure}

Therefore, Fig. 3 shows that, when ZrTe$_5$ undergoes a cyclic TPT ($Z_2$:1$\rightarrow$0$\rightarrow$1...) near $F_\text{th}\sim$170 $\mu$J/cm$^2$, $A_J$ (intensity of TRS-breaking), $A_\text{IR}$ (intensity of IS breaking), and $A_\text{charge}$ simultaneously reach their peaks. On the other hand, when $Z_2$:0$\rightarrow$0$\rightarrow$...or 1$\rightarrow$1$\rightarrow$..., $A_J$, $A_\text{IR}$, and $A_\text{charge}$ are apparently diminished. Particularly, this contrasts with, for instance, quantum Hall effects in terms of symmetry-topology interplay. In that case, by varying the magnetic field, the system transverses a series of topological states of different Chern numbers, and the system’s symmetry remains unchanged, i.e., symmetry is stable across TPT. In our case, however, the IR phonon mode (i.e., IS breaking) is generated and intensified at the TPT, even when the excitation frequency is far from the resonance of the IR phonon mode. The role of the light electric field is similar to the fluctuation in SSB in Landau theory and merely provides an initial perturbation to provoke the transition. Hence, the intensity of the IR phonon mode is not determined by the magnitude of the driving electric field, but by the extent of TPT which makes IS fragile. In addition, for the possible existence of SSB, we should point out that it is unsafe to presume that the crystal Hamiltonian $H$ may correctly reflect the resultant symmetry. For example, in our case $H$ perfectly respects both TRS and IS, which break due to the generation of photocurrents $J$ and IR phonons.

\begin{figure*}
	\begin{center}
		\includegraphics[width=160mm]{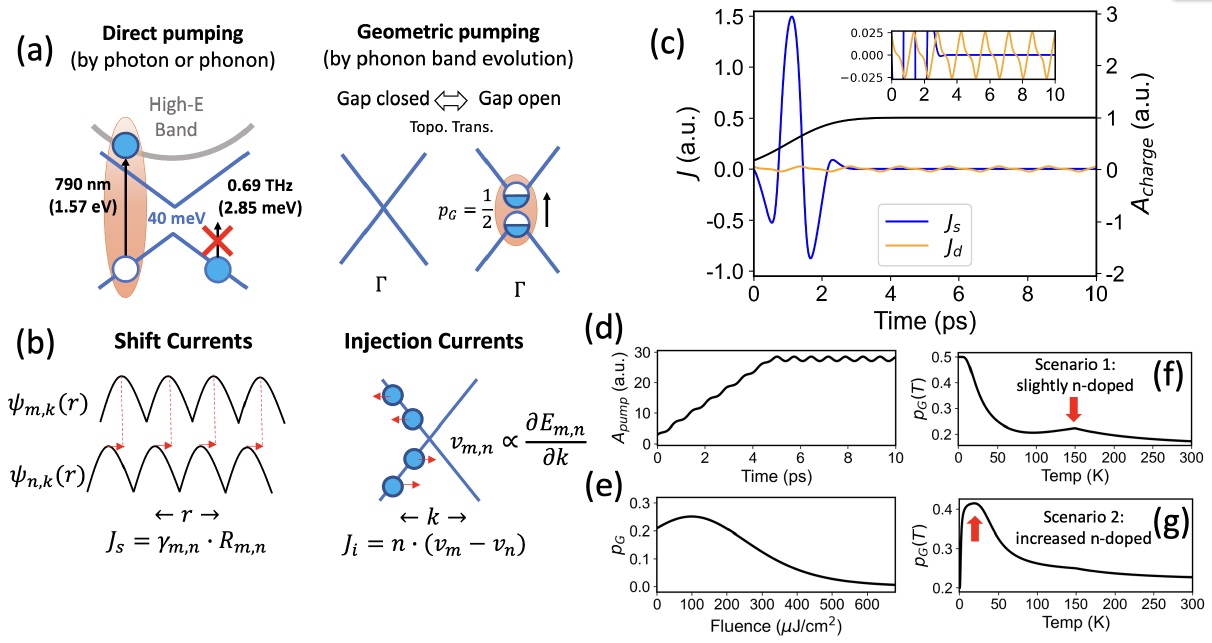}
		\caption{(a) Schematics to illustrate direct (energetic) pumping and geometric pumping. Geometric pumping occurs in gap-closing regions in the BZ with fractional pumping probability $p_G$=1/2 \cite{Ref42}, different from the normal direct pumping that requires energetic resonance and excites a “whole” quasi-particle. (b) Schematics for shift and injection currents. The former is due to Wannier center shift $R_{mn}$ \cite{Ref36,Ref37} when electrons hop between bands, and they reach the maximum when the hopping rate $\gamma$ is the largest. The latter is due to velocity difference $v_m-v_n$ across the two bands. For a linear dispersion, where velocity difference is fixed, the injection current $J_\text{i}$ relies on carrier density $n$ and thus peaks at the maximum density of $n$. Figs. 3 and 4 show that $J$ peaks at the maximum $\gamma$ (slope of charge pumping), indicating that the shift current $J_\text{s}$ dominates. (c) Theoretical calculation of time evolution of currents. The black curve denotes the charge pumping. (d) Simulated charge pumping process. (e) Fluence dependence of geometric pumping probability $p_G$. (f)(g) Temperature dependence of $p_G$ for two scenarios: slight $n$-doping and increased $n$-doping, respectively.}
		\label{Fig4}
	\end{center}
\end{figure*}

Another key evidence to underpin the critical role of TPT is to examine the three observables as a function of temperature.  
In Fig. 4(a), temperature-dependent THz signals at a fluence of $F$=40 $\mu$J/cm$^2$ are presented alongside the corresponding amplitude profile fittings. After subtracting the fitted amplitude profile from the THz signals, the resulting residual periodic oscillations for $t>$7 ps are Fourier transformed and plotted in Fig. 4(b), following the same methodology as in Figs. 3(a)-(b).
Here we fix the pump laser at a relatively low fluence, much smaller than $F_\text{th}$, so that the chemical potential is still located transiently above the Dirac point at $T$= 5 K as illustrated in the inset of Fig. 3(e). By increasing the temperature, the chemical potential will shift downwards, cross the cone vertex, and then continue to shift to below the Dirac point.
Temperature-dependent plots of $A_J$, $A_\text{IR}$, and $A_\text{charge}$ are displayed in Figs. 4(c)-4(e), respectively. 
Remarkably, the three distinct observables, namely photocurrents, IR phonons, and charge carrier density, exhibit a robust correlation with a shared temperature dependence. Similar to the fluence dependence, this shared behavior strongly indicates a common origin associated with the TPT.
Specifically, the amplitudes of all three exhibit a decline as temperature increases, with a notable upward cusp occurring around the Berry temperature, $T_\text{Berry}\sim$150 K. It is at this point that the TPT between the STI and WTI occurs. 
Notably, the charge pumping amplitude, $A_\text{charge}$, exhibits the most prominent bump at $T_\text{Berry}$, whereas $A_\text{IR}$ and $A_J$ show weaker ones.  
This observation aligns with our anticipated outcome of broken symmetry in the IR phonon and photocurrent near the TPT. At $T_\text{Berry}$, the conduction and valence bands come into contact at the $\Gamma$ point, leading to an enhancement in the transport and IR phonon observables. Next, we present a physical picture and model simulations to elucidate these observations.

\section{IV. Physical pictures and theoretical results}
The theme of the paper examines TPT acting as a trigger for (finite-lifetime) symmetry breaking, as observed from the behavior of electrons and phonons along the dimensions of time, fluence, and temperature. In this section, we present a theory that supports this by providing a mechanism that enables charge populations, transports, and lattices to be “hinged” together. Essentially, it reveals charge dynamics at TPT and uncovers an inter-band pumping mechanism: (i) this charge transfer shifts the charge center ${\langle}r{\rangle}$, driving charge transport (currents break TRS); (ii) the charge redistribution also distorts lattices (IR mode breaks IS). Thus, beyond its consistent agreement with experimental results (Figs. 2-4), the theory conveys a deeper insight: the “hinge” that links different aspects of the observations, which is independent of specific parameter details.

To be concrete, the theory is aimed to address three key facts from experiments: (1) The robust correlation among the three observables, $A_{J}$, $A_\text{IR}$, and $A_\text{charge}$, as a function of fluence and temperature. (2) The unusual decrease in their amplitudes as fluence exceeds $F_\text{th}$. (3) The enhancement of their amplitudes observed in proximity to the TPT temperature, $T_\text{Berry}$.

In Sec. IV. A, we account for the charge pumping $A_\text{charge}$. In Sec. IV. B, we address the two observables $A_{J}$ and $A_\text{IR}$, which are associated with TRS and IS breaking. In Sec. IV. C, we show the simulation (within a model that mimics ZrTe$_5$) of the three observables under varied fluences and temperatures to connect them with experimental data.

\subsection{A. Pumping mechanism}
We begin with charge pumping $A_\text{charge}$ mechanism. One might be tempted to consider that this pumping is directly excited by either photons (790 nm, 40 fs laser pulse) or IR phonons (0.69 THz). However, there is clear mismatch in time and energy scales among them. The duration of the charge pumping lasts for ${\sim} 7$ ps (red curve, Fig. 2(b)), apparently longer than the photon duration of ${\sim}$40 fs and shorter than the IR phonon duration of ${\sim}$tens of ps. In other words, during charge pumping $0<t<7$ ps, the material is in a field-free environment, and there is no photon to pump any charges. IR phonons can also be ruled out, because $A_\text{charge}$ should otherwise have been more long-lived and kept increasing until  IR phonons start to dissipate, i.e., $t\sim$tens of ps. From an energetic viewpoint, direct energetic pumping is unlikely, because the photons (1.57 eV) excite electrons to high-energy bands (Fig. 5(a)), which decay fast, irrelevant to the observed long-lasting, low-energy phenomena. The phonon's energy (2.85 meV) is evidently lower than the average band gap $\sim$40 meV. The mismatch for energy/time scales (summarized in Table I) together with abnormality in fluence/temperature dependence, alludes to unconventional pumping mechanism.

\begin{table}
	\caption{\label{tab:table1} Time and energy scales of the relevant phenomena. The time/energy scales of charge pumping and shift current $J_s$ (the two share common scales) are clearly distinct from applied photon (the first row) or phonon fields (the second row), indicating a non-energetic pumping mechanism.\label{tab1}}
	\begin{ruledtabular}
		\begin{tabular}{c c c}
			& Duration & Energy quanta \\
			\hline
			790 nm photon & 40 fs & 1.57 eV \\
			IR phonon & tens ps & 2.85 meV \\
			charge pumping \& $J_\text{s}$ & 5-10 ps & 40 meV (gap)
		\end{tabular}
	\end{ruledtabular}
\end{table}

Recently, a mechanism known as \textit{geometric pumping} has been unveiled, as illustrated in Fig. 5(a) \cite{Ref42, Ref42a}. It occurs when two bands touch at specific crystal momentum $k_0$ (either a point or a region), followed by a subsequent reopening of the gap (potentially driven by phonons, as referenced in \cite{Ref25,Ref29,Ref30,Ref31}). In this case, adiabaticity cannot be presumed, as gap closing ${\Delta}=0$ violates ${\omega}/{\Delta}{\rightarrow}0$. Physically, it means that even though phonons are comparatively slow compared to electrons, they behave as ``fast" near the gap closing. Thus at gap-closing point $k_0$, electrons potentially have non-zero pumping probability $p_G{\neq}0$.

In the study presented in \cite{Ref42}, an intricate analysis has been conducted, solving for the probability $p_G$. It leads to an elegant argument. If the closing of the gap results in a TPT, for instance, a cyclic transition of $Z_2$: 0 $\rightarrow$ 1 $\rightarrow$ 0, with each gap-closing flipping the topological index, then $p_G$=1/2. Conversely, if there is no TPT, $p_G$=0, effectively resembling an adiabatic evolution. Remarkably, the value of $p_G$ is solely contingent on the presence or absence of a TPT and remains insensitive to energetic details, such as phonon frequency or gap $\Delta$ (outside the closing regime).

To establish a connection with the realistic observables discussed in Section III, it is essential to consider the ensemble average to make a statistical argument. Because within the illuminated (macroscopic) area, there exists a multitude of subsystems, not all of which commence their evolution simultaneously. That means it requires finite time to establish $p_G$=1/2. We consider the time-resolving average of a collection of subsystems, 
\begin{equation}
A_\text{charge}(t)={\sum}_{i,n}p_G^{i,n}{\cdot}{\delta_\iota}(t-t_n^{i})
\end{equation} 
where $t_n^i$ stands for the $i^\text{th}$ subsystem's $n^\text{th}$ time of gap closing (TPT is cyclically driven by phonons). $p_G^{i,n}=((-1)^{n+1}+1)/2$ and $t_n^i=t_0^i+nT$. In the calculation, we adopt a Gaussian function of spreading ${\iota}=4$ ps to simulate the distribution of $t_0^i$. The larger $\iota$ is, the longer time it takes for every single subsystem to start vibration. The result is plotted in Fig. 5(d). Note that $\iota$ depends on factors, such as illuminated areas, laser conditions, etc. As a result, $\iota$ must be tuned to a ``plausible value" to match experiment. However, the non-tunability (constrained by the mechanism) is its linkage with transports (Fig. 5(c)). If $\iota$ was tuned to a different value, such as 8 ps, the main difference is simply that both Fig. 5(d) and 5(c) will get retarded correspondingly.

We emphasize several key points for interpreting the results: (i) The plateau observed in Fig. 5(d) corresponds to the system reaching a state of both conduction and valence bands being half-filled near $k_0$ ($\Gamma$ for ZrTe$_5$). (ii) The plateau state represents a dynamic equilibrium, in which evolution does not cease thereafter, while the occupancy reaches a stable point. Interestingly, as pointed out in Ref \cite{Ref42}, $p_G$=1/2 corresponds to the maximum entropy. (iii) Geometric pumping terminates when the phonon amplitude is too small to cause significant band distortion that can close the gap. Note that this turning point doesn't require the full relaxation of phonons. Afterwards, the accumulated charge in the upper band will gradually decay due to the lose of the pumping source, and this decay process should take an order of $\sim$ 100 ps as shown in Figs. 3(a) and 4(a). Therefore, the lifetime of geometric pumping ($\sim$ 7 ps) and the duration of excited charge carriers in the upper band ($\sim$100 ps) exhibit distinct notions. (iv) Phonons merely serve as a means to to realize TPT. {(v) The charge pumping dynamics persists for $\sim$7 ps, consistent with our prior study demonstrating that charge pumping occurs even with sub-gap excitation using low-energy THz pulses and the charge carrier population grew in step with coherent phonon oscillations rather than the pump pulse duration \cite{Ref25}, which provided compelling evidence for phonon-driven gap closing and highlighted charge carrier pumping as a versatile and robust indicator of gap closure during TPTs.


\subsection{B. Transport of excited charge carriers: shift currents.}
In this part, we consider the other two observables $A_J$ and $A_{\text{IR}}$, and their correlation with $A_{\text{charge}}$. There are two major transport mechanisms arising from optical pumping: shift current $J_\text{s}$ and injection current $J_\text{i}$ (Fig. 5(b)) \cite{Ref26, Ref36}. The former is sensitive to pumping rates, while the latter mainly depends on carrier density (given Fermi velocity is not vastly unchanged). In our case, the observed current $A_J$ peaks near the maximum slope of the charge population, i.e., proportional to the rate of pumping $\partial_{t}A_{\text{charge}}$. Therefore, it is most plausible to attribute the observed photocurrent $A_J$ to shift current $J_\text{s}$ associated with geometric pumping between the valence and conduction bands. Otherwise, if the current peaks at the maximum amplitude of the charge population $A_{\text{charge}}$, it is more likely to be the injection current $J_\text{i}$. As illustrated in Fig. 5(b), $J_\text{s}$ arises from a simple intuition: position shift when electrons are transferred $|{\psi}_m{\rangle}{\rightarrow}|{\psi}_n{\rangle}$. It is expressed as \cite{Ref26, Ref35, Ref36}
\begin{equation}
J_\text{s}=q{\cdot}{\gamma}_{m,n}{\cdot}R_{m,n}
\end{equation}
where $q$ is the charge of each carrier, $\gamma_{m,n}$ is the pumping rate $|{\psi}_m{\rangle}{\rightarrow}|{\psi}_n{\rangle}$ with the dimensionality [time]$^{-1}$, $R_{m,n}$ is the position shift between the two states. 
Pumping rate is linked with the derivative of charge pumping, 
\begin{equation}
{\gamma}_{m,n}~{\propto}~{\partial}_{t}A_\text{charge}(t).
\end{equation}
For the displacement vector $R_{m,n}$, it reads
\begin{equation}
R_{m,n}=A_{m,m}(k)-A_{n,n}(k)-{\frak{X}}_{m,n}(k)
\end{equation}
where $A_{m,m}$ is the Berry connection \cite{Ref26, Ref35, Ref36}, $m$ and $n$ are band labels; the complementary term ${\frak{X}}$ is to ensure the gauge invariant at a local $k$. 

If IS is respected, the integration of $R_{m,n}$ over the BZ will vanish, i.e., no net shift current \cite{Ref26}.
Therefore, phonons play a crucial role in this context, by driving the cyclic TPT, resulting in non-zero $p_G$ and ${\gamma}_{m,n}$. The cyclic TPT then leads to IS breaking, resulting in non-zero $R_{m,n}$. It is important to note that topology change and symmetry breaking are the fundamental physical mechanisms behind the generation of $J_\text{s}$, with phonons serving as a trigger of this process. In such a way, the IR phonon $A_\text{IR}$ is linked to the other two observables, $A_J$ and $A_\text{charge}$.

To understand the salient features of $J_\text{s}$, we consider a spinless two-band model $H(k)={\sum}_{i}d_i(k){\cdot}{\tau}_i$, where $d_i(k)$ is a real function about $k$, and ${\tau}_i$ is Pauli matrices for pseudo-spin. We choose $d_1(k)=v+\text{cos}(k)$, $d_2(k)=v'+\text{sin}(k)$, $d_3(k)=z$. In this model, the IS and TRS operators take the following forms \cite{Refa1}
\begin{equation}
I=I^{-1}={\tau}_1,~~\mathcal{T}=\mathcal{T}^{-1}=\mathcal{K}, 
\end{equation}
where $\mathcal{K}$ is a complex operator. It is easy to show that $v'$ will break both IS and TRS. The parameter $z$ is supposed to break IS, thus it should be linked to the instantaneous IR phonon amplitude $s(t)$, which could be expressed with $s(t)=s_0{\cdot}\text{sin}({\omega}t)$. With the model $H(k)$, one can evaluate Eq. (6), and by combining it with Eqs. (4)(5), $J_\text{s}$ can be calculated and plotted in Fig. 5(c). Note that Fig. 5(c) is intended for comparison with the red curve in Fig. 2(b). In the simulation, we neglected the trivial decay by scattering, thus the black curve in Fig. 5(c) ends up with a plateau without relaxation.


We emphasize a few key points: (i) the simulated results of $J_\text{s}$, $J_\text{d}$, and $A_\text{charge}$ shown in Fig. 5(c) reproduce the experimental results very well, as shown in Figs. 2(a)(b), 3(a) and 4(a). (ii) The displacement current $J_\text{d}$ due to bounded charges \cite{Refa1, Refa2} should be associated with the IR phonon mode. $J_\text{s}$ is due to interband hopping (geometric pumping), and $J_\text{d}$ is due to intraband (adiabatic) distortion, \cite{Refa1, Refa2}
\begin{equation}
J_\text{d}=\frac{e\dot{z}}{2\pi}{\int}{\langle}{\partial}_{z}u_{-}|i{\partial}_{k}u_{+}{\rangle}dk.
\end{equation}
(iii) During the first few ps when the charge pumping is taking place (the slope of the black curve in Fig. 5(c)), the interband contribution $J_\text{s}$ (blue in Fig. 5(c)) dominates. Subsequently when the charge pumping saturates, only intraband contribution $J_\text{d}$ (orange in Fig. 5(c)) remains. (iv) It is evident that $J_\text{s}{\gg}J_\text{d}$, which is not unexpected, because the current amplitude is determined by the ``hopping speed". For geometric pumping, in principle, it is infinitely fast, and generated immediately after the gap touches and reopens. In real systems, it corresponds to a short period of time, when the periodically oscillating bandgap becomes smaller than the phonon energy of ${\sim} 2.85$ meV. This time scale can be estimated by first-principle calculation to determine sub-gap regime in the midst of a whole phonon period, which turns out smaller than the phonon period by orders \cite{Ref24}. The intraband current $J_\text{d}$ is based on adiabatic distortion by IR phonons, which, however, have the same time scale as the IR phonon period. Thus, $J_{\text{d}}$ is much smaller. (v) The real physical causes of $J_\text{s}$ is TPT and TRS breaking, and phonons merely trigger to realize the two conditions. Therefore, that's the reason we stated that $J_\text{s}$ is induced by phonons (or by laser pulses), but not directly driven by it. Because, apparently $J_\text{s}$ is \textit{not} always proportional to the phonon amplitude (and laser intensity), but the intensity of TPT, and it only requires $|s(t)|$ reaching a threshold to activate gap closing. Accordingly, the decay of $|J_\text{s}|$ is due to the decay of the geometric pumping rate $\gamma~{\propto}~{\partial}_{t}A_{\text{charge}}{\rightarrow}0$ rather than the vanishing of phonons. (vi) The observable $J_s$ reflects the intensity of TRS breaking driven by TPT, which should not be confused with phonon-field-driven current in analog with $E$-field-driven current, which can be expressed as $J_\text{s}={\sum}_{n}{\chi}^{(n)}{\cdot}E^n$ \cite{Ref35, Ref36}. Since phonon-field $s(t)$ only needs to reach a threshold, therefore a polynomial expansion shown below is invalid,
\begin{equation}
J_\text{s}{\neq}{\chi}^{(1)}{\cdot}s+{\chi}^{(2)}{\cdot}s^{2}+...
\end{equation}

In this context, it is necessary to further distinguish the meanings of “induced by” and “driven by”. Basically, the shift currents $J_{\text{s}}$ are induced (or triggered) by laser electric fields but \textit{not} directly driven by the fields. For currents driven by fields (either electric field or phonon field), they are defined by Eq. (9), where conductivities $\chi$ should be positive. Thus, the bottom line is that the magnitude of field-driven currents positively depend on the amplitude of fields whether it follows a linear or higher-order fashion. Physically, that means the current is driven by the force from the fields. On the other hand, ``induced by" or ``triggered by" means the external field is necessary for current generation but will not directly determine the magnitude of currents. It is similar to the role of fluctuation in phase transitions: it is fluctuation that triggers the phase transition, while the magnitude of formed order parameter is determined by the free energy. In other words, fluctuation merely pushes the system to provide an “initial momentum”, but irrelevant to where the system will end up with. In this case, the current  is only triggered by electric field, whose function is merely to trigger the TPT; while the magnitude of currents depends on the extent of TPT.

\subsection{C. Simulations of fluence and temperature dependence}
The theoretical approach above is obtained for $T$=0 K with the Fermion occupancy $f_v=1$ and $f_c=0$. At finite temperatures, there are four possible situations as outlined in Table II, 
\begin{table}
	\caption{\label{tab:table1} Occurrence or non-occurrence of geometric pumping in a given occupancy situation. $f_v$ and $f_c$ refer to the Fermion distribution of valence and conduction bands, respectively. occ. and emp. refer to occupied and empty states, respectively. \label{tab1}}
	\begin{ruledtabular}
		\begin{tabular}{c c c c c}
			valence & conduction & Geometric pumping & Probability & $p_G$ \\
			\hline
			occ. & occ. & no & $f_vf_c$ & 0 \\
			occ. & emp. & yes & $f_v(1-f_c)$ & 1/2 \\
			emp. & occ. & yes & $(1-f_v)f_c$ & 1/2 \\
			emp. & emp. & no & $(1-f_v)(1-f_c)$ & 0
		\end{tabular}
	\end{ruledtabular}
\end{table}
where the pumping probability is
\begin{equation}
0{\cdot}f_vf_c+\frac{1}{2}{\cdot}f_v(1-f_c)+\frac{1}{2}{\cdot}(1-f_v)f_c+0{\cdot}(1-f_v)(1-f_c).
\end{equation}
This leads to a unique temperature feature of geometric pumping
\begin{eqnarray}  \label{equ3}
\begin{aligned}
p_G(T)=(f_\nu+f_c-2f_{\nu}f_c)\cdot p_G(T=0).
\end{aligned}
\end{eqnarray}
This arises from the fact that geometric pumping requires one band being occupied and the other being empty \cite{Ref42,Refa3,Refa4}. Eq. \ref{equ3} is clearly distinct from an energetic transition such as described by the Fermi-Golden rule, which states that $p(T)=(f_v-f_c)\cdot p(T$=0). In that case, if valence and conduction bands switch, i.e., $c\leftrightarrow \nu$, thus $f_v-f_c \rightarrow-(f_v-f_c$)  reverses the sign. The sign reversal could be interpreted as the existence of an energetic preference direction. In contrast, Eq. \ref{equ3} will preserve the sign, thus it has no preference direction, which is an important distinctive signature for geometric pumping. 

Eq. \ref{equ3} is also applicable to the fluence dependence, because both fluence- and temperature-dependence fundamentally involves tuning the position of the chemical potential $\mu$ with respect to the Dirac point. We use the most essential linear response model (i.e., ${\mu}(F)$ and ${\mu}(T)$ are linear functions) to capture the key major features of $p_G$. Results are shown in Figs. 5(e)-(g). The fluence dependence of $p_G$ (Fig. 5(e)) exhibits a good agreement with the experimental results in Figs. 3(c)-(e), and the temperature dependence (Figs. 5(f)(g)) is consistent with the experimental results in Figs. 4(c)-(e).

We should revisit the key experimental features we try to capture as raised at the beginning of this section. The fluence dependent $p_G$ exhibits an initial increase followed by a reduction as shown in Fig. 5(e). This is because the sample is initially $n$-doped \cite{Ref26} and the pump beam can transiently move $\mu$ downwards as discussed in Section III. The maximum of $p_G$ corresponds to $\mu$ crossing the cone vertex, occurring at a threshold fluence. Above the threshold fluence, $\mu$ moves further downwards resulting in empty conduction and valence bands at the $\Gamma$ point, which leads to $p_G{\rightarrow}0$. For the temperature dependent $p_G$ shown in Figs. 5(f)(g), the simulation shows that it is sensitive to the $n$-doping level. There are two scenarios: if there is only a small amount of $n$-doping, $p_G$ exhibits the maximum at $T$=0 K and a small bump at $T_{\text{Berry}}\sim$150 K. However, as the doping level increases, $p_G$ at $T$=0 K drops quickly due to both conduction and valence bands being occupied at the $\Gamma$ point. Additionally, the peak at $T_{\text{Berry}}$ tends to be smoothed. Comparing our simulations to the experimental data in Fig. 4, we find that the system resides in an intermediate state between the two scenarios. This observation is entirely consistent with the presence of a finite yet small Fermi energy, approximately 15 meV, as determined in our sample at 5 K. It is indeed intriguing that our model based on a coarse linear $\mu$-dependence manages to capture not only the general trends, but also the small bump near $T_{\text{Berry}}$. This finding underscores the robustness of our conclusions regarding the geometric pumping mechanism. The linear temperature dependence of energy levels for the two scenarios is shown in Supplemental Fig. 1. 

\section{V. Discussion and Outlook}
This work performs multiple-dimensional observations, and the essential points could be summarized in several aspects:

(1)	Fluence: All the three observables ($A_J$, $A_{\text{IR}}$, $A_{\text{charge}}$) first increase then decrease with fluence, displaying a similar dependence throughout the range, whether it is near the peak$\sim$170 $\mu$J/cm$^2$ or far away.

(2)	Temperature: All the three observables display a similar trend in the temperature domain, especially a kink around 150 K, where temperature-driven TPT happens.

(3) Time: Firstly, there is clear mismatch among the charge pumping time scale ($\sim$7 ps), the IR mode lifetime ($\sim$tens of ps), and laser pulse duration ($\sim$40 fs), which indicates the charge pumping is neither primarily caused by fields nor by IR phonons, consistent with the scenario of being caused by geometric pumping at TPT \cite{Ref42}. Secondly, there is a clear correlation between the current $J_{\text{s}}$ and the slope of $A_{\text{charge}}$, i.e., $J_{\text{s}}$ is proportional to ${\partial}_tA_{\text{charge}}$.

(4)	Energy: The three hinged observables feature quite distinct energy scales, as outlined in Table I.

The model connects these observations. First, it provides a mechanism that correlates the three observables $A_J$, $A_{\text{IR}}$, $A_{\text{charge}}$. It establishes a physical scenario that shift currents may only depend on TPT and TRS breaking, instead of being driven by laser fields.
Second, it justifies the abnormality of the fluence dependence results shown in Fig. 3. The photocurrents are not light electric field-driven, thus $J_\text{s}{\propto}E^n$ is invalid, i.e., $J_\text{s}$ does not monotonically increase with $E$. The primary impact of the laser pulses is to tune the chemical potential and its relative position with respect to the conical point at $\Gamma$. This adjustment in energy levels directly influences the occupancy and subsequently affects the intensity of TPT and geometric pumping. The effect is most pronounced when the TPT is most intense, i.e., when the chemical potential $\mu$ crosses the band cone vertex or when conduction and valence bands touch. Thirdly, in the temperature dependence, our model predicts that the signal for all the three observables should exhibit a peak at TPT, i.e., $T_{\text{Berry}} \sim$ 150 K. Furthermore, it anticipates a pronounced sensitivity to $n$-doping levels, aligning with the small Fermi energy measurements obtained in our samples.

The symmetry breaking discovered here is reminiscent of the Jahn-Teller effect (JTE) \cite{Ref43}, for which molecules will distort at electronic degeneracy. In that case, JTE applies to molecular or ligand systems. The structural transition of molecules is triggered by the degeneracy of electronic levels, and the degeneracy is lifted after the transition. It is about a permanent distortion at a given temperature, which breaks spatial symmetries \cite{Ref43}. In our case, however, it is a crystal system that is studied, in which the triggering condition is TPT, instead of energetic degeneracy, and the broken symmetry is about temporal generation of the IR phonon mode (there is not a symmetry breaking throughout the temperature range we have examined) and it involves TRS, not just spatial group. The symmetry instability is triggered by a dynamic transition of topological states.


We also rule out light-induced strain effects which arise from non-uniformity (fluence field $F(r)$) within the light spot. The magnitude is proportional to gradient $|{\nabla}F(r)|$. Typically, increasing pump fluence leads to a larger gradient $|{\nabla}F(r)|$. Therefore, the decreased signals with fluence contradict such scenarios. Furthermore, it is difficult to conceive how strain effects could simultaneously influence all the three observables in a correlated manner.

A shortcoming of the current theory is that it does not provide a mutual-feedback mechanism between electron and lattices. Thus, a fair description could be that the present work provides a plausible scenario of tuning TPT. Generally speaking, TPT is much harder to detect and control than symmetry, especially in ultrafast processes, where the key signatures, edge states, become difficult to catch. This is a broad challenge. This work has taken several ways in tackling the challenge. First, we compared signals from ZrTe$_5$ with non-topological materials, such as InAs as shown in Fig. 3(f) and GaAs \cite{DCoteAPL2002}. Second, we cross-check with different dimensions, such as fluence- and temperature-driven TPT. Third, we turn to theoretical clues for assistance. Fortunately, ref \cite{Ref42} indicates that this phenomenon is prevalent in topological insulators and semimetals, and more supporting evidence is expected in the near future. Our findings highlight the need for in-depth exploration of electron-phonon coherence, leveraging advanced THz 2D coherent spectroscopy \cite{2D1, 2D2} and THz near-field microscopy \cite{nano1, nano2}.




\section{VI. CONCLUSION}
In this study, we present a compelling demonstration within a model system showcasing the profound impact of topology on symmetries, ultimately leading to spontaneous symmetry breaking.
We explored the breaking of TRS and IS by observing photocurrent generation and IR phonons in the vicinity of TPT by discovering an intriguing anomaly in the fluence and temperature dependence measurements. 
Contrary to conventional expectations driven by direct, energetic excitations, we observe that the extent of symmetry breaking does not exhibit a straightforward
positive correlation with laser fluence. Instead, our results reveal a complex interplay between the TPT and the observables $A_J$, $A_\text{IR}$, and $A_\text{charge}$.
These compelling insights suggest a profound connection between the broken symmetry observables and the occurrence of TPTs, where they synergistically enhance each other. Our work pushes the boundaries of our understanding of topology and symmetry, uncovering a captivating intersection between them. 


\section{ACKNOWLEDGEMENTS}

This work was supported by the Ames National Laboratory, the US Department of Energy, Office of Science, Basic Energy Sciences, Materials Science and Engineering Division under contract No. DE-AC02-07CH11358 (Ultrafast and THz photocurrent spectroscopy and model building). Sample development and magneto-transport measurements in Brookhaven National Laboratory (Q.L., P.M.L., G.G.) were supported by the US Department of Energy, Office of Basic Energy Science, Materials Sciences and Engineering Division, under contract No. DE-SC0012704.

\bibliographystyle{apsrev}

\begin{thebibliography}{99}
\bibitem{Ref38} For current-driven field, one can expand it in terms of $J(\omega)=\sigma_1 E(\omega)+\sigma_2 E(\omega)(-\omega)+...$. In this case, electric field disappears after 40 fs. Thus, field-driven currents (to arbitrary orders) should vanish quickly after electric field is gone, and the expansion formula does not apply here.
\end{thebibliography}

\end{document}